\documentclass[runningheads]{llncs}
\usepackage{graphicx,xcolor}
\usepackage{url}
\newcommand{\numFactors}{11~}
\usepackage{multirow}
\begin{document}
\title{Should I visit this place? Inclusion and Exclusion Phrase Mining from Reviews}
\titlerunning{Inclusion and Exclusion Phrase Mining from Reviews}

\newcommand{\floatValue}{8pt}
\renewcommand\textfloatsep{\floatValue}
\renewcommand\floatsep{\floatValue}
\renewcommand\intextsep{\floatValue}
\renewcommand\dbltextfloatsep{\floatValue}
\renewcommand\dblfloatsep{\floatValue}

\author{Omkar Gurjar$^1$ \and Manish Gupta$^{1,2}$}
%
\authorrunning{Gurjar and Gupta}
\institute{$^1$IIIT-Hyderabad, India\ \ \ \  $^2$Microsoft, India \\
\email{omkar.gurjar@students.iiit.ac.in, manish.gupta@iiit.ac.in, gmanish@microsoft.com}}
%
\maketitle 
\begin{abstract}
Although several automatic itinerary generation services have made travel planning easy, often times travellers find themselves in unique situations where they cannot make the best out of their trip. Visitors differ in terms of many factors such as suffering from a disability, being of a particular dietary preference, travelling with a toddler, etc. While most tourist spots are universal, others may not be inclusive for all. In this paper, we focus on the problem of mining inclusion and exclusion phrases associated with \numFactors such factors, from reviews related to a tourist spot. While existing work on tourism data mining mainly focuses on structured extraction of trip related information, personalized sentiment analysis, and automatic itinerary generation, to the best of our knowledge this is the first work on inclusion/exclusion phrase mining from tourism reviews. Using a dataset of 2000 reviews related to 1000 tourist spots, our broad level classifier provides a binary overlap F1 of $\sim$80 and $\sim$82 to classify a phrase as inclusion or exclusion respectively. Further, our inclusion/exclusion classifier provides an F1 of $\sim$98 and $\sim$97 for 11-class inclusion and exclusion classification respectively. We believe that our work can significantly improve the quality of an automatic itinerary generation service.
\end{abstract}

\section{Introduction}
\label{sec:intro}
Hundreds of millions of visitors travel across the globe every year resulting into trillions of dollars of spending. Number of international tourist arrivals has seen a steady increase over the past few decades\footnote{\url{https://data.worldbank.org/indicator/ST.INT.ARVL}}. Thanks to the availability of multiple online services like web maps, travel and stay booking, and automatic planning, tourism has become a lot comfortable in recent years. 

Automated itinerary planning systems\footnote{\url{http://itineree.com/top-online-travel-planners/}} provide a holistic solution enabling transportation, lodging, sights, and food recommendations. 
However such recommendation systems cannot incorporate subtle user constraints like a claustrophobic user, visitors travelling with a toddler, visitors of a particular ethnicity with visa restrictions, etc. Indeed, many of them, do not even incorporate tourist spot specific properties like what time of day is best to visit, temporary ad hoc closures due to local vacations or maintenance work, visitor height/gender restrictions, vegetarian friendly or not, etc.  

Tourist review websites are a gold mine of data related to very subtle restrictions (or exclusions) associated with a tourist spot. In this work, we focus on the following 11 different factors regarding inclusion or exclusion nature of tourist spots. 
(1)Age/Height: Disallow visitors of a particular age/height group: too old, or too young, too short.
(2) Claustrophobia: Some spots consist of a lot of confined spaces and hence unsuitable for claustrophobic visitors.
(3) Couples/Family: Some spots are family/kids friendly versus not.
(4) Crowd: Some spots are often heavily crowded, which may be repulsive to some visitors.
(5) Food: Some spots may serve low quality food, non-vegetarian food only, may not serve any food, may not allow any external food, may not allow alcoholic drinks, etc.
(6) Handicap: Some spots may not allow facilities for disabled folks like lifts, ramps, etc. The terrain may not be wheel-chair or stroller friendly.
(7) Hygiene: Some spots may be filthy, e.g., unclean toilets, littered beaches, etc. 
(8) Parking: Unavailability and ease of parking.
(9) Price: Some spots may be very expensive for tourists.
(10) Queues: Some spots may exhibit large queues leading to long wait times. Visitors on a tight schedule may want to avoid such places, or visit them in low wait time durations.
(11) Time: Various spots have a preferred visit timings, such as early morning, late evening, on Wednesdays, from Sep-Dec, etc. This category also includes ad hoc closures due to maintenance or other reasons. 

In this paper, we focus on two related tasks: (1) Task 1 pertains to mining inclusion/exclusion phrases from tourism reviews. A phrase which pertains to any of the exclusions as mentioned above is labeled as an exclusion phrase, while a phrase related to inclusion of the above factors is labeled as an inclusion phrase. (2) Task 2 is about fine-grained classification of inclusion/exclusion phrases into one of the above 11 categories. ``I had my kids who loved this museum'' and ``elevators for those whom stairs are problematic'' are examples of age and handicap inclusion phrases. ``place was very crowded'', ``would not recommend the area for young children'' are examples of crowd and age exclusion phrases.

We are the first to propose the problem of extracting inclusion/exclusion phrases from tourism review data. 
The problem is challenging: (1) There can be many types of exclusions as discussed above. (2) These factors can be expressed in lots of different ways. (3) There could be multiple indirect references (e.g. if the place allows gambling, likely kids are not allowed) or unrelated references (e.g., a review talking about a tour guide's ``family'' rather than if ``families'' are allowed at the spot).


Overall, we make the following contributions in this paper: (1) We propose a novel task of mining inclusion/exclusion phrases from online tourist reviews, and their fine-grained classification. 
(2) We model the first task as a sequence labeling problem, and the second one as a multi-class classification. We investigate the effectiveness of CRFs (Conditional Random Fields), BiLSTMs (Bidirectional Long Short-Term Memory networks), and Transformer models like BERT (Bidirectional Encoder Representations from Transformers).
(3) We make the code and the manually labeled dataset (2303 phrases mined from $\sim$2000 reviews) publicly available\footnote{\url{https://github.com/omkar2810/Inclusion_Exclusion_Phrase_Mining}}. Our experiments show that the proposed models lead to practically usable classifiers.



\section{Related Work}
\label{sec:relatedWork}
\noindent\textbf{Tourism Data Mining}: 
Work on tourism data mining has mostly focused on structured extraction of trip related information~\cite{popescu2009deducing}, mining reviews (personalized sentiment analysis of tourist reviews~\cite{pantano2017you}, establishing review credibility~\cite{ayeh2013we,filieri2015travelers}), and automatic itinerary generation~\cite{chang2016atips,chen2013automatic,de2010automatic,friggstad2018orienteering}. Popescu et al.~\cite{popescu2009deducing} extract visit durations or information like ``what can I visit in one day in this city?'' from Flickr data. Pantano et al.~\cite{pantano2017you} predict tourists' future preferences from reviews. Ayeh et al.~\cite{ayeh2013we} examine the credibility perceptions and online travelers' attitude towards using user-generated content (UGC). Filieri et al.~\cite{filieri2015travelers} study the impact of source credibility, information quality, website quality, customer satisfaction, user experience on users'  trust towards UGC. The automatic itinerary generation problem has been studied extensively from multiple perspectives. Friggstad et al.~\cite{friggstad2018orienteering} model the problem as an orienteering problem on a graph of tourist spots. Chang et al.~\cite{chang2016atips} weigh different factors like spot name, popularity, isRestaurant, isAccomodation, etc. based on user interactions to optimize the process of trip planning. De et al.~\cite{de2010automatic} aggregate across geo-temporal breadcrumbs data for multiple users to construct itineraries. 
Clearly, our system can be an important sub-module to generate automated itineraries which are exclusion-sensitive.

\noindent\textbf{Sequence Labeling}: 
Sequence labeling involves predicting an output label sequence given an input text sequence. A label is generated per input token. Popular sequence labeling models include CRFs~\cite{lafferty2001conditional}, LSTMs~\cite{graves2013hybrid}, LSTM-CRFs~\cite{huang2015bidirectional}, and Transformer models like BERT~\cite{devlin2018bert}. Many NLP tasks can be modeled as sequence labeling tasks including opinion mining~\cite{irsoy2014opinion}, part-of-speech tagging, etc. The labels for such tasks are typically encoded using BIO (begin, inside, outside) labeling. In this paper, we investigate the effectiveness of such sequence labeling approaches for the inclusion/exclusion phrase mining task.

\noindent\textbf{Aspect Extraction}: Aspect extraction has been studied widely in the past decade, mainly for product reviews, using supervised~\cite{wang2016recursive}, semi-supervised~\cite{mukherjee2012aspect} as well as unsupervised~\cite{he2017unsupervised} methods. In this work, we study aspect extraction for reviews in the tourism domain.

\section{Proposed Approach}
\label{sec:approach}
\subsection{Dataset}
We first obtained a list of top 1000 tourist spots from lonelyplanet.com (a popular tourist website). Next, we obtained a maximum of 2000 reviews corresponding to each of these spots from tripadvisor.com. Further, we broadly filtered out reviews (and then sentences) that could be potentially related to the eleven factors mentioned in Sec.~\ref{sec:intro} using a manually produced keyword list for each category. We provide the full keyword list per category as part of the dataset. These $\sim$2000 reviews were then manually labeled for inclusion/exclusion phrases using the BIO tagging, as well as their fine categorization into one of the 11 categories. A total of 2303 phrases were labeled with one of the 11 categories. The distribution across the categories is as follows: Age/Height: 324, Claustrophobia: 217, Couples/Family: 151, Crowd: 307, Food: 313, Handicap: 204, Hygiene: 95, Parking: 65, Price: 351, Queues: 185, and Time: 91. For the inclusion/exclusion phrase mining task, a total of 2303 phrases from 2154 sentences were labeled. Phrases in these sentences which are not inclusion/exclusion are marked as others. Across these phrases, the word label distribution is as follows: B\_EXC: 1176, B\_INC: 1223, EXC: 5713, INC: 5455, O: 29976, where INC and EXC denote inclusion and exclusion respectively. We make the code and the manually labeled dataset publicly available$^3$. On a small set of 115 instances, we measured the inter-annotator agreement and found the Cohen's Kappa to be 0.804 and 0.931 for the first and the second tasks respectively, which is considered as very good.
\subsection{Methods}
We experiment with two different word embedding methods: GloVe (Global Vectors for Word Representation)~\cite{pennington2014GloVe} and ELMo (Embeddings from Language Models)~\cite{peters2018deep}. We use CRFs, BiLSTMs, BiLSTM-CRFs and BERT for the first sequence labeling task. We use traditional machine learning (ML) classifiers like XGBoost and Support Vector Machines (SVMs) and deep learning (DL) models like BiLSTMs, LSTM-CNN and BERT for the multi-class classification task.

\noindent\textbf{CRFs~\cite{lafferty2001conditional}}: Conditional Random Fields (CRFs) are prediction models for tasks where contextual information or state of the neighbors affect the current prediction. They are a type of discriminative undirected probabilistic graphical model. 

\noindent\textbf{BiLSTMs~\cite{graves2013hybrid}}: Bidirectional LSTMs are the most popular traditional deep learning models for sequence modeling. They model text sequences using recurrence and gate-controlled explicit memory logic. Bidirectionality helps propagate information across both directions leading to improved accuracies compared to unidirectional LSTMs. 

\noindent\textbf{BiLSTM-CNNs~\cite{chiu2016named}}: BiLSTM-CNNs use character-based CNNs to first generate the word embeddings. These word embeddings are further used by the LSTM to generate the embedding for the text sequence. This is then connected to a dense layer and then finally to the output softmax layer. 

\noindent\textbf{BiLSTM-CRFs~\cite{huang2015bidirectional}}: We combine a BiLSTM network and a CRF network to form a BiLSTM-CRF model. This network can efficiently use past input features via a LSTM layer and sentence level tag information via a CRF layer. 

\noindent\textbf{BERT~\cite{devlin2018bert}}: BERT is a Transformer-encoder model trained in a bidirectional way. BERT has been shown to provide very high accuracies across a large number of NLP tasks. For the sequence labeling task, we connect the semantic output for each position to an output softmax layer. For multi-class classification, we connect semantic representation of CLS token to the output softmax layer. 

\section{Experiments}
\label{sec:experiments}
For BiLSTM experiments, we used three layers, ReLU activation for hidden layers and softmax for output, SGD optimizer (with momentum=0.7, learning rate=1e-5, batch size=8), and cross-entropy loss. We trained for 50 epochs. We used GloVe 200D word vectors. For BERT, we used the pretrained BERT BASE model with 12 Transformer layers, Adam optimizer with learning rate=3e-5, max sequence length=128, batch size=8, and categorical cross entropy loss. 

\subsection{Results}
Table~\ref{tab:level1Results} shows results for the inclusion/exclusion phrase mining  task. As discussed in~\cite{irsoy2014opinion}, we use two metrics: (1) Binary Overlap which counts every overlapping match between a predicted and true expression as correct, and (2) Proportional Overlap which imparts a partial correctness, proportional to the overlapping amount, to each match. BERT based method outperforms all other methods. This is because the 12 layers of self-attention help significantly in discovering the right inclusion/exclusion label for each word. Also, precision values are typically lower than recall, which means that our models can detect that the text implies some inclusion or exclusion but find it difficult to differentiate between the two.

\begin{table}
    \centering
        \scriptsize 
    \begin{tabular}{|p{0.9in}|c|c|c|c|c|c|c|c|c|c|c|c|}
\hline
\multirow{3}{*}{Model}&\multicolumn{6}{c|}{Inclusion}&\multicolumn{6}{c|}{Exclusion}\\
\cline{2-13}
&\multicolumn{2}{c|}{Precision}&\multicolumn{2}{c|}{Recall}&\multicolumn{2}{c|}{F1}&\multicolumn{2}{c|}{Precision}&\multicolumn{2}{c|}{Recall}&\multicolumn{2}{c|}{F1}\\
\cline{2-13}
&Prop&Bin&Prop&Bin&Prop&Bin&Prop&Bin&Prop&Bin&Prop&Bin\\
\hline
CRF + GloVe&0.354&0.417&0.531&0.758&0.425&0.538&0.372&0.392&0.524&0.728&0.435&0.512\\
\hline
BiLSTM + GloVe&0.456&0.590&0.573&0.643&0.508&0.615&0.506&0.638&0.570&0.668&0.536&0.650\\
\hline
BiLSTM CRF + GloVe&0.490&0.625&0.613&0.714&0.545&0.666&0.516&0.649&0.654&0.788&0.577&0.712\\
\hline
BiLSTM + ELMo&0.580&0.645&0.604&0.770&0.590&0.701&0.602&0.678&0.566&0.738&0.579&0.703\\
\hline
BERT&\textbf{0.677}&\textbf{0.748}&\textbf{0.765}&\textbf{0.869}&\textbf{0.718}&\textbf{0.804}&\textbf{0.664}&\textbf{0.756}&\textbf{0.801}&\textbf{0.908}&\textbf{0.726}&\textbf{0.825}\\
\hline
    \end{tabular}
    \caption{Inclusion/Exclusion Phrase Mining Accuracy Results}
    \label{tab:level1Results}
\end{table}

We present the results of our 11-class phrase classification in Table~\ref{tab:level2Results}. We observe that typically the accuracy is better for inclusion phrases rather than exclusion phrases. Deep learning based methods like LSTMs and BERT are better than traditional ML classifiers. BERT outperforms all other methods by a large margin for both the inclusion and exclusion phrases.    

Further, we performed an end-to-end evaluation of our system. For each sentence in the test set, we first obtained BIO predictions using our phrase mining system. Then, we perform 11-class classification on these mined phrases. Golden label for our predicted inclusion/exclusion phrase is set to the ground truth label for the phrase with maximum intersection. For predicted phrases which have no intersection with any golden phrase, we assume them to belong to a special ``sink'' class, and they count towards loss in precision. Golden phrases not detected by our system count towards loss in recall. Such an evaluation leads to an overall F1 of 0.748 (P=0.695, R=0.812), inclusion F1 of 0.739 (P=0.691, R=0.795) and an exclusion F1 of 0.759 (P=0.700, R=0.830).
\begin{table}
    \centering
    \scriptsize
    \begin{tabular}{|l|c|c|c|c|c|c|c|c|c|c|}
    \hline
\multirow{2}{*}{Model}&\multicolumn{3}{c|}{Total}&\multicolumn{3}{c|}{Inclusion}&\multicolumn{3}{c|}{Exclusion}\\
    \cline{2-10}
    &Precision&Recall&F1&Precision&Recall&F1&Precision&Recall&F1\\
   \hline
 SVM&0.725&0.631&0.626&0.759&0.635&0.649&0.665&0.626&0.604\\
   \hline
 XGBoost&0.802&0.796&0.797&0.802&0.785&0.786&0.817&0.806&0.806\\
   \hline
 BiLSTM + GloVe&0.890&0.885&0.884&0.921&0.917&0.916&0.862&0.852&0.853\\
   \hline
 BiLSTM-CNN + GloVe&0.895&0.892&0.891&0.903&0.900&0.900&0.889&0.883&0.883\\
   \hline
 BiLSTM Attn + GloVe&0.914&0.911&0.911&0.938&0.934&0.934&0.894&0.887&0.889\\
   \hline
 BERT&\textbf{0.978}&\textbf{0.978}&\textbf{0.978}&\textbf{0.983}&\textbf{0.982}&\textbf{0.982}&\textbf{0.975}&\textbf{0.973}&\textbf{0.973}\\
 \hline
    \end{tabular}
    \caption{11-class Categorization Accuracy Results}
    \label{tab:level2Results}
\end{table}

Next, we present two examples of the output from our system. Consider the sentence: ``The wheelchair wouldn't go through the turnstile which was disappointing''. Our inclusion/exclusion phrase mining BERT classifier outputs ``B\_EXC EXC EXC EXC EXC EXC EXC O O O'' while our 11-class classifier labels this as ``Handicap''. Our system was able to smartly associate ``wheelchair wouldn't go through'' with ``handicap'' category. Consider another example, ``We came to Eiffel Tower to celebrate twenty five years of togetherness''. Our two classifiers predict ``O O O O O O O INC INC INC INC INC'' and ``Couples/Family''. Interestingly, it can relate ``togetherness'' with ``Couples/Family''.

\subsection{Error Analysis}
We performed a manual analysis of some of the errors made by our best model. We found the following interesting patterns. (1) It is difficult to predict the right label when the phrase can be provided multiple labels. E.g. ``If you don't like crowds or feel claustrophobic being on narrow walkways full of groups of people ...'' can be labeled into either of the Crowd or Claustrophobia categories. (2) Conflicting opinions mentioned in same review. ``... Well worth the \$25 ... The cost of the day was very expensive compared to Australian water parks.'' In this review, from a price perspective, it is difficult to figure out whether the spot is cheap or expensive. Similarly, consider another review: ``Wednesday night is bike night in Beale Street so a lot of noise from at least 1000 bikes many highly decorated. It was fun and the usual bar street of many cities.'' Can't really make out whether one should visit during the night or not. (3) References to other unrelated things: Consider this review: ``... I was lucky enough to have a descendant who gave the garden tour and tell about the family (more than you might usually get) ...'' The word ``family'' here does not indicate anything about inclusion/exclusion wrt families for the spot. 

\section{Conclusion}
\label{sec:conclusion}
In this paper, we proposed a novel task for mining of inclusion/exclusion phrases and their detailed categorization. We investigated the effectiveness of various deep learning methods for the task. We found that BERT based methods lead to a binary overlap F1 of $\sim$80 and $\sim$82 for the sequence labeling task, and an F1 of $\sim$98 and $\sim$97 for 11-class inclusion and exclusion classification respectively. In the future, we plan to integrate this module as a part of a personalized automated itinerary recommendation system.
 \bibliographystyle{splncs04}
 \bibliography{references}

\begin{thebibliography}{10}
\providecommand{\url}[1]{\texttt{#1}}
\providecommand{\urlprefix}{URL }
\providecommand{\doi}[1]{https://doi.org/#1}

\bibitem{ayeh2013we}
Ayeh, J.K., Au, N., Law, R.: {``Do we believe in TripAdvisor?'' Examining
  credibility perceptions and online travelers’ attitude toward using
  user-generated content}. Journal of Travel Research  \textbf{52}(4),
  437--452 (2013)

\bibitem{chang2016atips}
Chang, H.T., Chang, Y.M., Tsai, M.T.: {ATIPS: automatic travel itinerary
  planning system for domestic areas}. Computational intelligence and
  neuroscience  (2016)

\bibitem{chen2013automatic}
Chen, G., Wu, S., Zhou, J., Tung, A.K.: Automatic itinerary planning for
  traveling services. IEEE transactions on knowledge and data engineering
  \textbf{26}(3),  514--527 (2013)

\bibitem{chiu2016named}
Chiu, J.P., Nichols, E.: Named entity recognition with bidirectional lstm-cnns.
  Transactions of the Association for Computational Linguistics  \textbf{4},
  357--370 (2016)

\bibitem{de2010automatic}
De~Choudhury, M., Feldman, M., Amer-Yahia, S., Golbandi, N., Lempel, R., Yu,
  C.: Automatic construction of travel itineraries using social breadcrumbs.
  In: Proceedings of the 21st ACM conference on Hypertext and hypermedia. pp.
  35--44 (2010)

\bibitem{devlin2018bert}
Devlin, J., Chang, M.W., Lee, K., Toutanova, K.: Bert: Pre-training of deep
  bidirectional transformers for language understanding. arXiv preprint
  arXiv:1810.04805  (2018)

\bibitem{filieri2015travelers}
Filieri, R., Alguezaui, S., McLeay, F.: Why do travelers trust tripadvisor?
  antecedents of trust towards consumer-generated media and its influence on
  recommendation adoption and word of mouth. Tourism management  \textbf{51},
  174--185 (2015)

\bibitem{friggstad2018orienteering}
Friggstad, Z., Gollapudi, S., Kollias, K., Sarlos, T., Swamy, C., Tomkins, A.:
  Orienteering algorithms for generating travel itineraries. In: Proceedings of
  the Eleventh ACM International Conference on Web Search and Data Mining. pp.
  180--188 (2018)

\bibitem{graves2013hybrid}
Graves, A., Jaitly, N., Mohamed, A.r.: Hybrid speech recognition with deep
  bidirectional lstm. In: 2013 IEEE workshop on automatic speech recognition
  and understanding. pp. 273--278. IEEE (2013)

\bibitem{he2017unsupervised}
He, R., Lee, W.S., Ng, H.T., Dahlmeier, D.: An unsupervised neural attention
  model for aspect extraction. In: Proceedings of the 55th Annual Meeting of
  the Association for Computational Linguistics (Volume 1: Long Papers). pp.
  388--397 (2017)

\bibitem{huang2015bidirectional}
Huang, Z., Xu, W., Yu, K.: Bidirectional lstm-crf models for sequence tagging.
  arXiv preprint arXiv:1508.01991  (2015)

\bibitem{irsoy2014opinion}
Irsoy, O., Cardie, C.: Opinion mining with deep recurrent neural networks. In:
  Proceedings of the 2014 conference on empirical methods in natural language
  processing (EMNLP). pp. 720--728 (2014)

\bibitem{lafferty2001conditional}
Lafferty, J.D., McCallum, A., Pereira, F.C.: Conditional random fields:
  Probabilistic models for segmenting and labeling sequence data. In:
  Proceedings of the Eighteenth International Conference on Machine Learning.
  pp. 282--289 (2001)

\bibitem{mukherjee2012aspect}
Mukherjee, A., Liu, B.: Aspect extraction through semi-supervised modeling. In:
  Proceedings of the 50th Annual Meeting of the Association for Computational
  Linguistics (Volume 1: Long Papers). pp. 339--348 (2012)

\bibitem{pantano2017you}
Pantano, E., Priporas, C.V., Stylos, N.: ‘you will like it!’ using open
  data to predict tourists' response to a tourist attraction. Tourism
  Management  \textbf{60},  430--438 (2017)

\bibitem{pennington2014GloVe}
Pennington, J., Socher, R., Manning, C.D.: Glove: Global vectors for word
  representation. In: Proceedings of the 2014 conference on empirical methods
  in natural language processing (EMNLP). pp. 1532--1543 (2014)

\bibitem{peters2018deep}
Peters, M.E., Neumann, M., Iyyer, M., Gardner, M., Clark, C., Lee, K.,
  Zettlemoyer, L.: Deep contextualized word representations. arXiv preprint
  arXiv:1802.05365  (2018)

\bibitem{popescu2009deducing}
Popescu, A., Grefenstette, G.: Deducing trip related information from flickr.
  In: Proceedings of the 18th international conference on World wide web. pp.
  1183--1184 (2009)

\bibitem{wang2016recursive}
Wang, W., Pan, S.J., Dahlmeier, D., Xiao, X.: Recursive neural conditional
  random fields for aspect-based sentiment analysis. arXiv preprint
  arXiv:1603.06679  (2016)

\end{thebibliography}
\end{document}